\documentstyle[aps,prb,preprint]{revtex} 
\begin{document}
\title{ \LARGE Electronic structure study of double perovskites
$A_{2}$FeReO$_{6}$ ($A$=Ba,Sr,Ca) and Sr$_{2}$$M$MoO$_{6}$ 
($M$=Cr,Mn,Fe,Co) by LSDA and LSDA+$U$} 

\author{ Hua Wu\\}

\address{\normalsize
 Max-Planck-Institut f\"ur Physik komplexer Systeme, D-01187 Dresden, Germany\\
 and Institute of Solid State Physics, Academia Sinica,
           230031 Hefei, P. R. China\\}

\maketitle
\begin{abstract}

  We have implemented a systematic LSDA and LSDA+$U$ study of the double 
perovskites
$A_{2}$FeReO$_{6}$ ($A$=Ba,Sr,Ca) and Sr$_{2}$$M$MoO$_{6}$
($M$=Cr,Mn,Fe,Co) for understanding of their intriguing electronic and magnetic
properties. The results suggest a ferrimagnetic (FiM) and half-metallic (HM)
state of $A_{2}$FeReO$_{6}$ ($A$=Ba,Sr) 
due to a $pdd$-$\pi$ coupling between the down-spin
Re$^{5+}$/Fe$^{3+}$ $t_{2g}$ orbitals via the intermediate O $2p_{\pi}$ ones, 
also a very similar FiM and HM state
of Sr$_{2}$FeMoO$_{6}$. In contrast, a decreasing Fe 
$t_{2g}$ component at Fermi level ($E_{F}$) in the distorted 
Ca$_{2}$FeReO$_{6}$ partly accounts for its nonmetallic behavior, while
a finite $pdd$-$\sigma$ coupling between the down-spin Re$^{5+}$/Fe$^{3+}$ 
$e_{g}$
orbitals being present at $E_{F}$ serves to stabilize its FiM state.
For Sr$_{2}$CrMoO$_{6}$ compared with Sr$_{2}$FeMoO$_{6}$, the 
coupling between the down-spin 
Mo$^{5+}$/Cr$^{3+}$ $t_{2g}$ orbitals decreases as a noticeable shift up
of the Cr$^{3+}$ $3d$ levels, which is likely responsible for the decreasing 
$T_{\rm C}$ value and weak conductivity.
Moreover, the calculated level distributions indicate a  
Mn$^{2+}$(Co$^{2+}$)/Mo$^{6+}$ ionic state in Sr$_{2}$MnMoO$_{6}$
(Sr$_{2}$CoMoO$_{6}$), in terms of which their antiferromagnetic 
insulating ground state
can be interpreted. While orbital population analyses show that owing to
strong intrinsic $pd$ covalence effects, Sr$_{2}$$M$MoO$_{6}$
($M$=Cr,Mn,Fe,Co) have nearly the same valence state combinations, as accounts 
for the similar $M$-independent spectral features observed in them.  \\

\end{abstract}

\newpage

\section*{I. INTRODUCTION}

  A recent finding of the room-temperature tunnelling 
magnetoresistance (TMR) effect in double perovskites 
Sr$_{2}$FeMoO$_{6}$ (SFMO)\cite{c1}
and Sr$_{2}$FeReO$_{6}$ (SFRO)\cite{c2} revives the study of $3d$ ($M$) and 
$4d$/$5d$
($M'$) transition-metal oxide alloys $A_{2}MM'$O$_{6}$ ($A$=Ba,Sr,Ca).
Most of the alloys have been up to now found to take a rock-salt crystal 
structure, an ordered one of the alternate perovskite units $AM$O$_{3}$ and
$AM'$O$_{3}$ along three crystallographical axes.\cite{c3} SFMO
and SFRO taking this structure are predicted by band calculations
to be a HM,\cite{c1,c2} where energy bands of one spin channel (up or down) 
cross $E_{F}$
and are therefore metallic but those of the other spin channel are separated by 
an insulating gap. Theoretically speaking, the conduction electrons are 100$\%$
spin polarized, this effect generally related to a ferromagnetic (FM) ordering
of a (sub)lattice, despite a possible inter-sublattice antiferromagnetic (AFM)
coupling in some 
cases. For the polycrystallic ceramics of SFMO and SFRO,
an applied magnetic field can control magnetic domain reorientation
and tends to parallelize the magnetization directions of FM domains and 
therefore reduces the spin scattering of spin-polarized carriers at grain 
boundaries, giving rise to a significant decrease of the measured 
resistivity.\cite{c1,c2}
Such an effect was referred to as an interdomain/grain TMR.  
Like the well-known HM-like colossal MR manganites,\cite{c4} the TMR materials 
also have
potential technological applications to magnetic memory and actuators. 

The MR materials with an interesting combination of electronic and magnetic 
properties are of current considerable interest also from the basic points of
view. Double exchange (DE)\cite{c5} model has been widely employed to account 
for the
FM metallicity of La$_{1-x}$Ca$_{x}$MnO$_{3}$ and related compounds,
in spite of its recent modification by taking into account a strong 
electron-phonon coupling arising from the Jahn-Teller splitting of 
the Mn$^{3+}$ ions,\cite{c6} and even an alternative $p$-$d$ exchange model 
as lately suggested for those O $2p$-$M$ $3d$ charge-transfer oxides.\cite{c7}
One DE-like mechanism\cite{c8,c9} was suggested for SFMO [SFRO] 
that a strong hybridization between the Mo$^{5+}$ 
($4d^{1}$) [Re$^{5+}$ ($5d^{2}$)] and Fe$^{3+}$ ($3d^{5}$)
down-spin $t_{2g}$ orbitals via a $pdd$-$\pi$ coupling leads to electron 
mobility being responsible for the HM behavior, and 
the itinerant down-spin carriers are
antiferromagnetically polarized by the localized Fe$^{3+}$ $S$=5/2 full 
up-spins and therefore mediate a FM coupling in the Fe sublattice. 
Alternatively, very late a new mechanism\cite{c9,c10} was proposed that if the 
Fe sublattice
has a FM ordering (being so actually), a bonding-antibonding splitting due to
the Mo $4d$ [Re $5d$]/Fe $3d$ hybridization results in a shift up (down)  
of the Mo $4d$ [Re $5d$] $t_{2g}^{\uparrow}$ ($t_{2g}^{\downarrow}$) bands
located between the Fe$^{3+}$ $3d$ full-filled $t_{2g}^{\uparrow}$ and
empty $t_{2g}^{\downarrow}$ bands, and therefore an electron transfer 
from Mo [Re] $t_{2g}^{\uparrow}$ bands to $t_{2g}^{\downarrow}$ ones.   
This energy gain causes a negative spin polarization of the formally 
nonmagnetic Mo [Re] species, and the itinerant electrons in the partly 
filled $t_{2g}^{\downarrow}$ conduction bands cause kinetic energy gain further
via the DE interaction, both of which commonly stabilize the FM state,
as compared with the AFM or paramagnetic (PM) state.

 In contrast, it was argued in another late study\cite{c11} that a direct 
Re $t_{2g}$-Re $t_{2g}$ interaction is the main cause for the metallic behavior
of Ba$_{2}$FeReO$_{6}$ (BFRO), and a lattice distortion of Ca$_{2}$FeReO$_{6}$
(CFRO)
induced by the smaller-size Ca species disrupts the Re-Re interaction and makes
itself nonmetallic. While magnetization measurements show that the latter 
has a higher $T_{\rm C}$ value than the former, which seems surprising when
referring to their distinct conduction behaviors.\cite{c11}

One of the aims of this work is to probe which mechanism,
either the direct or the indirect Re-Re interaction is essentially 
responsible for the conduction behavior of BFRO and SFRO by means of 
density functional theory (DFT)\cite{c12} calculations. The present results 
clearly show that the Re-O-Fe-O-Re interaction rather than the direct Re-Re one
takes the responsibility. And it is 
suggested, based on the calculated orbital-resolved density of states
(DOS), that a reduced Re $t_{2g}$-Fe $t_{2g}$ $pdd$-$\pi$ hybridization in 
distorted CFRO
partly accounts for its nonmetallic behavior, compared with
the cases of cubic BFRO and SFRO. While
a presence of a finite Re $e_{g}$-Fe $e_{g}$ $pdd$-$\sigma$ hybridization in 
CFRO at $E_{F}$
contributes to its increasing $T_{\rm C}$. 

On the other hand,
the dependence of the electronic, magnetic, and transport properties of
Sr$_{2}M$MoO$_{6}$ ($M$=Cr,Mn,Fe,Co) on the $3d$ transition-metal species
is an intriguing issue.\cite{c8,c13} 
SFMO and Sr$_{2}$CrMoO$_{6}$ have a high $T_{\rm C}$ value, and the former is 
a metallic compound while the latter is a nonmetal with a relatively low
room-temperature resistivity.\cite{c8} In contrast, Sr$_{2}$MnMoO$_{6}$ and
Sr$_{2}$CoMoO$_{6}$ are AFM insulators.\cite{c13}
For the reason, DFT calculations are also 
designed for them. The valence states, spin moments, electronic and magnetic 
characters are discussed below in detail.

\section*{II. Computational details}
 
  The structure data of $A_{2}$FeReO$_{6}$ ($A$=Ba,Sr,Ca) and 
Sr$_{2}M$MoO$_{6}$ ($M$=Cr,Mn,Fe,Co) are taken from Refs. 11, 2 and 8.
These compounds have a cubic crystal structure except for the tetragonal
Sr$_{2}$CoMoO$_{6}$ and monoclinic CFRO. 
An orthogonal cell is assumed for CFRO in order to
simplify calculations, since the experimentally determined monoclinic cell
($\beta$=90.02) deviates slightly from an orthogonal one.\cite{c11} 
FM calculations are performed for these compounds. As seen below, there is
an induced negative spin polarization of the Re or Mo sublattice at the 
background of the FM Fe and/or Cr sublattices, which we call a FiM structure
in this paper. In addition, the present FM solutions of  
Sr$_{2}$MnMoO$_{6}$ and Sr$_{2}$CoMoO$_{6}$  
can be extended to an explanation of their AFM 
insulating behaviors.  

  The full-potential linearly combined atomic-orbital (LCAO) band 
method,\cite{c14} 
based on the local spin density approximation (LSDA)\cite{c12} to DFT and
on-site Coulomb correlation correction (LSDA+$U$)\cite{c15}, is adopted in
the present calculations. 
Hartree potential is expanded in terms of lattice harmonics up to $L$=6, and an 
exchange-correlation potential of von Barth-Hedin type\cite{c16} is adopted.
The $U$ (=3, 4, 4.5, 5 eV) parameters are used for the strongly 
correlated $M$ (=Cr,Mn,Fe,Co) $3d$ electrons,\cite{c17} respectively; while a 
small $U$=1 eV for the weakly correlated Re $5d$ and Mo $4d$ 
electrons.\cite{c18} Ba $5p5d6s$/Sr $4p4d5s$/Ca $3p3d4s$, $M$ $3d4s$, 
Re $5d6s$/Mo $4d5s$, and O $2s2p$ orbitals are treated as valence states. 
125 (64 for CFRO with a doubled cell) special $k$ points in 
irreducible Brillouin 
zone are used in the present self-consistent calculations.

  A description of the method for orbital population analyses is given below 
for the reason that a detailed discussion about the orbital occupation
is made in the text.

  A crystal wave function $\Psi_{j}^{\bf k}$ is expressed in terms of
Bloch basis functions \{$\Phi_{l}^{\bf k}$\} in the LCAO formalism:
\begin{equation}
\Psi_{j}^{\bf k}=\sum_{l}C_{l}^{{\bf k}j}\Phi_{l}^{\bf k}=\sum_{l}
C_{l}^{{\bf k}j}\frac{1}{\sqrt{N}}\sum_{m}e^{i{\bf k}\cdot(\tau+
{\bf R}_{m})}\phi_{l}({\bf r}-\tau-{\bf R}_{m)},
\end{equation}
where $\phi_{l}$ denotes an atomic orbital wave function. ${\bf k}$, $j$, $l$, 
$N$, ${\bf R}_{m}$, and $\tau$
stand for, respectively, a wave vector, a band index, an atomic orbital label, 
number of unit cells, a lattice vector, and an atomic position.

  The total DOS reads
\begin{eqnarray}
\rho(E)&=&\frac{1}{N_{\bf k}}\sum_{{\bf k}j}\left|\Psi_{j}^{\bf k}\right|^{2}
\delta(E-E_{j}^{\bf k})=\frac{1}{N_{\bf k}}\sum_{{\bf k}j}\sum_{ll'}
C_{l}^{{\bf k}j}C_{l'}^{{\bf k}j*}S_{l'l}^{{\bf k}}
\delta(E-E_{j}^{\bf k}) \nonumber\\
&=&\frac{1}{N_{\bf k}N}\sum_{{\bf k}j}\sum_{lml'm'}C_{l}^{{\bf k}j}
C_{l'}^{{\bf k}j*}<\phi_{l'}|\phi_{l}>e^{i{\bf k}\cdot
(\tau+{\bf R}_{m}-\tau'-{\bf R}_{m'})}\delta(E-E_{j}^{\bf k}),
\end{eqnarray}
where
\begin{equation} 
S_{l'l}^{{\bf k}}=<\Phi_{l'}^{{\bf k}}|\Phi_{l}^{{\bf k}}> 
\end{equation}
refers to 
an overlap matrix, and 
$N_{\bf k}$ represents the number of ${\bf k}$-points. 

  $\rho(E)$ can be decomposed into an on-site term $\rho_{A}(E)$ 
($m$=$m'$ and $\tau$=$\tau'$) and an overlap term $\rho_{B}(E)$
($m$$\neq$$m'$ or $\tau$$\neq$$\tau'$). The on-site term is written as
\begin{equation}
\rho_{A}(E)=\frac{1}{N_{\bf k}}\sum_{{\bf k}j}\sum_{l}
\left|C_{l}^{{\bf k}j}\right|^{2}
\delta(E-E_{j}^{\bf k})
\end{equation}
due to the orthogonality and normality of atomic orbitals at the same site.
Correspondingly, the on-site occupation number $Q_{A}$
\begin{equation}
Q_{A}=\int\rho_{A}(E)dE=\frac{1}{N_{\bf k}}\sum_{{\bf k}j}\sum_{l}
\left|C_{l}^{{\bf k}j}\right|^{2}
\end{equation}
is composed of atomic contributions \{$\left|C_{l}^{{\bf k}j}\right|^{2}$\}.
Obviously, the sum rule for total number of electrons is not satisfied due to 
the non-orthogonality of
atomic orbitals at different sites. For the reason, the overlap term 
$\rho_{B}(E)$ is dealt with by Mulliken analysis\cite{c19} which is widely
used among quantum chemistry community and provides a reasonable
description of local orbitals. 
 
  The overlap term is expressed as 
\begin{eqnarray}
\rho_{B}(E)&=&\frac{1}{N_{\bf k}}\sum_{{\bf k}j}\sum_{l,l'\neq l}
C_{l}^{{\bf k}j}C_{l'}^{{\bf k}j*}S_{l'l}^{{\bf k}}\delta(E-E_{j}^{\bf k})
\nonumber \\
&=&\frac{1}{N_{\bf k}}\sum_{{\bf k}j}\sum_{l}\sum_{l'\neq l}
\frac{\left|C_{l}^{{\bf k}j}\right|^{2}}{\left|C_{l}^{{\bf k}j}\right|^{2}
+\left|C_{l'}^{{\bf k}j}\right|^{2}}(C_{l}^{{\bf k}j}C_{l'}^{{\bf k}j*}
S_{l'l}^{{\bf k}}+C_{l}^{{\bf k}j*}C_{l'}^{{\bf k}j}S_{ll'}^{{\bf k}})
\delta(E-E_{j}^{\bf k}) \nonumber \\
&=&\frac{1}{N_{\bf k}}\sum_{{\bf k}j}\sum_{l}
D_{l}^{{\bf k}j}\delta(E-E_{j}^{\bf k}).
\end{eqnarray}
Mulliken analysis is taken such way that the overlap term is decomposed into
atomiclike contributions $D_{l}^{{\bf k}j}$, according to their respective 
`weight' $\left|C_{l}^{{\bf k}j}\right|^{2}$. 

  As a result, one can calculate the orbital-resolved DOS and population,
respectively, by the expressions  
\begin{equation}
\rho(E)=\rho_{A}(E)+\rho_{B}(E)=\frac{1}{N_{\bf k}}\sum_{{\bf k}j}\sum_{l}
(\left|C_{l}^{{\bf k}j}\right|^{2}+D_{l}^{{\bf k}j})\delta(E-E_{j}^{\bf k}),
\end{equation}
\begin{equation}
Q=\frac{1}{N_{\bf k}}\sum_{{\bf k}j}\sum_{l}
(\left|C_{l}^{{\bf k}j}\right|^{2}+D_{l}^{{\bf k}j}).
\end{equation}
Note that for a spin-polarized system, one can calculate the contributions
from up- and down-spin channels and then obtain the values of spin moments.

\section*{III. Results and Discussions}

  The LSDA calculations for BFRO and SFRO 
give a nearly same HM solution as recently reported for
the latter.\cite{c2,c10} Their level distributions are overall very similar,
as is not surprising since, on one hand, both compounds have the same
cubic structure except for a difference in their lattice
constants 8.05 \AA~ versus 7.89 \AA~;\cite{c11,c2} on the other hand, two 
Ba$^{2+}$/Sr$^{2+}$ ions per formula unit donate their four valence electrons,
[all $A$(=Ba,Sr,Ca) almost taking a valence state of +1.8 and a minor spin
moment of --0.01 $\mu_{B}$ in the present calculations], into 
(FeReO$_{6}$)$^{-4}$ unit and scarcely affect the valence bond 
interactions both in and between the (FeReO$_{6}$)$^{-4}$ units.
Owing to the lattice distortion of CFRO---shrinked Fe-O-Re
bond lengths and bent bond angles\cite{c11}---and therefore modified 
$pdd$ hybridizations, however, a noticeable difference emerging in 
CFRO is that the down-spin $e_{g}$ hybridized states are also present
at $E_{F}$. It can be seen in Fig. 1 that in $A_{2}$FeReO$_{6}$ ($A$=Ba,Sr,Ca),
the up-spin
Fe $3d$ orbitals are almost full-filled, while the down-spin Fe $3d$ ones are
nearly empty except for a small amount of occupations due to $pd$ 
hybridizations, both of which indicate that the formal Fe$^{3+}$ ($3d^{5}$) 
ions actually take a mixing $3d^{5}$+$3d^{6}L$ state ($L$: a ligand hole) like
an ordinary case, as also seen in Table I. Moreover, 
a small increase of Fe $3d$ population (and a corresponding minor decrease of
Fe spin) as the change of $A$ from Ca via Sr to Ba, being
qualitatively in accord with the M\"{o}ssbauer spectroscopy study,\cite{c11} 
indicates a decreasing oxidization of Fe, which is also supported by the
calculated valence state of the coordinated oxygens, --1.47 for $A$=Ca,
--1.4 for $A$=Sr, and --1.35 for $A$=Ba. 
While a significantly strong $pd$ hybridization effect is evident for
the formal high valence Re$^{5+}$ ions, which causes a larger 
bonding-antibonding splitting and a crystal-field $t_{2g}$-$e_{g}$ splitting.   
As a result, although the down-spin Re $t_{2g}$ states crossing $E_{F}$ are
partly occupied as the expected 2/3 filling, the $5d$ orbital population
is larger than 4 and far away from the formal Re$^{5+}$ $5d^{2}$ as seen in
Table I. A similar population difference larger than 2 also appears in the 
high valence oxides, $e.$$g.$, V$_{2}$O$_{5}$\cite{c20} and 
NaV$_{2}$O$_{5}$.\cite{c21} 
The relatively smaller Re $5d$ population in CFRO, compared with SFRO and BFRO,
is also related to the relatively higher negative valence of the coordinated
oxygens. A stronger ionicity of CFRO could be implied by a weaker Fe-O-Re
covalence interaction caused by the bent Fe-O-Re bond angle. Moreover, 
a little smaller Re $5d$ population in SFRO than in BFRO could be 
due to a more delocalized behavior of the down-spin Re $t_{2g}$ orbitals 
in SFRO, which is closely related to a larger band width caused by the smaller
cubic lattice constant of SFRO.  

 As seen in Fig. 1, the Re $t_{2g}$ levels exactly lie between 
the up- and down-spin Fe $t_{2g}$ ones, and they are close to the
down-spin Fe $t_{2g}$ levels. As a result, 
the bonding-antibonding
mechanism stated above induces a spin splitting of the Re $5d$ levels
and a subsequent electron transfer and therefore a negative spin polarization.
Moreover, it is evident that a stronger Re/Fe bonding-antibonding interaction
occurs in the down-spin channel than in the up-spin one.
The present LSDA calculations give the Fe (Re) spin moments of 3.72 
(--0.99) $\mu_{B}$ in BFRO, and 3.75 (--1.05) $\mu_{B}$ in SFRO, and 
3.76 (--1.07) 
$\mu_{B}$ in CFRO, all of which are reduced by the $pd$ hybridizations,
compared with the ideal 5 (--2) $\mu_{B}$ for Fe$^{3+}$ (Re$^{5+}$). 
A stronger Re $5d$ spin polarization in SFRO than in BFRO could be due to 
a stronger bonding-antibonding mechanism in SFRO caused by the shorter Fe-O-Re
bond. While the calculated largest Re spin in CFRO is ascribed to both
an additional negative spin polarization of the Re $e_{g}$ electrons and weaker
delocalization of the Re $5d$ electrons.
The calculated total spin, including both the weakly polarized oxygen spins
with an averaged value of 0.05 $\mu_{B}$ and the minor $A$ spin, is accurately 
equal to 3 $\mu_{B}$ per formula unit as an expected integral value 
for a HM. While the reduced experimental values, $e.$$g$. 2.7 $\mu_{B}$
in Sr$_{2}$FeReO$_{6}$,\cite{c2} are probably ascribed to
a site-disorder effect as previously suggested.\cite{c1,c22}   
The calculated Fe spin moments agree well with the previous 
3.7 $\mu_{B}$ in SFRO,\cite{c10} while the Re spin moments are larger 
than the corresponding value of --0.78 $\mu_{B}$.\cite{c10} 
Although the previous values were calculated within relatively small radii 
($e.$$g.$, muffin-tin spheres);\cite{c10} while the present ones are calculated
for the magnetic
ions with extended orbitals, it is not surprising that the Fe moments (nearly
75$\%$ spin-polarized) are
almost identical because of the strong localization of the Fe $3d$ orbitals.
The present larger Re moments (about 50$\%$ spin-polarized) are 
ascribed to the delocalized Re $5d$ orbital behavior. 
   
  A high Re-O (and Fe-O) covalent charge density is evident in Fig. 2. 
In addition, the up-spin Fe charge density is nearly spherical, which 
corresponds to the formal full-filled up-spin Fe$^{3+}$ $3d$
orbital; while the down-spin Re (and Fe) density is obviously of the 
$t_{2g}$ orbital-like 
distribution. Moreover, the Re-Re interstitial densities visually arising from
the O $2p$ contribution almost have no difference between the up- and down-spin
channels, which indicates that the net
down-spin Re charge scarcely serves to the Re-Re interaction. In other word,
the direct Re-Re interaction due to the itinerant down-spin Re $t_{2g}$ 
electrons is nearly impossible. Furthermore, the charge densities in 
the Re-O-Fe bond are quite larger than
the Re-Re interstitial densities. A combination of these results with the DOS 
shown in Fig. 1 leads to a suggestion that the down-spin 
Re $t_{2g}$-O $2p_{\pi}$-Fe $t_{2g}$ $pdd$-$\pi$ coupling rather than the 
direct Re-Re interaction is responsible for 
the FiM and HM character of BFRO and SFRO. 

 In the calculated FiM and HM state of CFRO, however, the down-spin
Fe $t_{2g}$ component decreases at $E_{F}$ due to bent Fe-O-Re bond angles.
The reduced Fe-O-Re $pdd$-$\pi$ coupling, as well as a little stronger 
site-disorder effect than in BFRO,\cite{c11} could partly account for the 
nonmetallicity 
of CFRO. While a finite down-spin Re $e_{g}$-O $2p_{\sigma}$-Fe $e_{g}$ 
$pdd$-$\sigma$ 
coupling emerges at $E_{F}$ in CFRO, which is due to the $t_{2g}$-$e_{g}$ 
mixing caused by the lattice distortion. The presence of such a coupling could 
stabilize the FiM state of CFRO and 
leads to its increasing $T_{\rm C}$ as compared with BFRO.\cite{c11} 
However, the reasons for the nonmetallicity but high $T_{\rm C}$ of CFRO 
are not yet definitely clear.
 
The $U$ correction lowers the occupied up-spin Fe $3d$ bands significantly 
and makes them much narrower and thus behave like a localized $S$=5/2 spin.
The O $2p$ states undergo a noticeable change, while the Re $5d$ bands
change insignificantly due to a weak electron correlation. 
The present LSDA+$U$ 
results of SFRO are well comparable with those lately 
reported with a close value $U_{eff}$=4 eV for Fe.\cite{c10} The HM character 
remains 
unchanged, and particularly, the DOS at $E_{F}$ [$N(E_{F})$] caused mostly
by the down-spin Re $t_{2g}$-O $2p$ hybridized states keeps the 
value of about 4 states/eV per formula unit (nearly twice as large as that of 
SFMO\cite{c1,c10}) as the above LSDA result, since the 
Re species are not strongly correlated ones. The LSDA+$U$ calculations
increase the Fe (Re) spin moments up to 4.16 (--1.22) $\mu_{B}$ in BFRO,
4.22 (--1.27) $\mu_{B}$ in SFRO, and 4.29 (--1.33) $\mu_{B}$ in CFRO. The 
increases of the Fe spin moments are nearly twice as large as those of the Re
moments, due to a stronger localization and electron correlation of the Fe ions.
In addition, the increasing difference between the present and previous 
LSDA+$U$ Re spin moments is ascribed to the $U$=1 eV correction to the Re $5d$
states which is adopted in the present calculations but not in previous 
ones.\cite{c10}

Now we turn to Sr$_{2}$$M$MoO$_{6}$
($M$=Cr,Mn,Fe,Co). First, the LSDA results of SFMO (see Fig. 3)
are in good agreement with the previous ones.\cite{c1,c10} Like the above case 
of SFRO,
the present Fe spin moment of 3.78 $\mu_{B}$ is almost identical with 
two independently calculated values of 3.79\cite{c1} and 3.73 
$\mu_{B}$,\cite{c10} 
and the Re spin moment of --0.44 $\mu_{B}$ is larger than two 
corresponding ones of --0.29\cite{c1} and --0.30 $\mu_{B}$.\cite{c10}  
These Re moments 
are all less spin-polarized ($<$45$\%$). These results suggest once again that 
the Fe spin is strongly localized while the Mo spin is delocalized. 
In SFMO the 
itinerant down-spin Mo $t_{2g}$ electrons mediate a FM coupling between the full
Fe up-spins and are responsible for its HM character via the Mo-O-Fe coupling,
like the above cases of BFRO and SFRO.
Owing to the $4d^{1}$ configuration of the formal Mo$^{5+}$ ions,  
$N(E_{F})$ ($\approx$2 states/eV per formula unit being consistent with
the previous results\cite{c1,c10}) is reduced by nearly one 
half, compared with the above cases of BFRO and SFRO with Re$^{5+}$ $5d^{2}$.
It can be seen in Fig. 3 that besides a partial filling of the down-spin Mo 
$t_{2g}$ orbitals (formally 1/3 occupied), there is a quite large
amount of Mo $4d$ occupations between --8 and --4 eV due to strong $pd$ 
covalence effects. The Mo ions actually have a configuration nearly $4d^{3.4}$ 
(see Table I).

For the case of Sr$_{2}$CrMoO$_{6}$, owing to a shift up of the Cr$^{3+}$ $3d$ 
levels compared
with the Fe$^{3+}$ case of SFMO, the down-spin Mo $t_{2g}$-O $p_{\pi}$-Fe 
$t_{2g}$
coupling is reduced around $E_{F}$, which could account for its decreasing 
$T_{\rm C}$ and nonmetallic behavior\cite{c8} (the latter as discussed 
above 
for CFRO). Besides, the occupied up-spin Cr$^{3+}$ $t_{2g}$ levels are very 
close to $E_{F}$ and near the empty up-spin Mo$^{5+}$ $t_{2g}$ ones. Thus the 
enhanced hybridization
between them (also due to the relatively short Cr-O-Mo bond) leads
to a broadened bandwidth, which makes the up-spin $t_{2g}$ hybridized bands 
crossing $E_{F}$ also and partly occupied. The calculated Cr (Mo) spin moment 
is 1.67 (--0.65) $\mu_{B}$, and the total spin moment per formula unit is 1.13 
$\mu_{B}$, which is increased by the following $U$ correction up to the 
expected 2 $\mu_{B}$ for an ideal FiM state. While  
an experimental magnetization value of $\approx$0.5 $\mu_{B}$\cite{c8} is much 
smaller
than these theoretical ones. A significant site-disorder effect could account 
for this large discrepancy,\cite{c1,c22} since a minor difference between the 
ionic radii (Cr$^{3+}$ 0.615 \AA~, Mo$^{5+}$ 0.61 \AA~)\cite{c23} most probably
leads to
a random structure.\cite{c3} If a pair of antisite Cr/Mo ions leads to a 
decrease of
spin moments by 6 $\mu_{B}$, according to the analyses in Ref. 22, the maximal 
antisite portion is estimated to be 25$\%$. Note that the formal Mo$^{5+}$ ions
take $4d^{3.34}$ in the LSDA calculation. The Mo spin is larger in
Sr$_{2}$CrMoO$_{6}$ than in SFMO, reflecting a weaker itineracy of the 
Mo $t_{2g}$ electrons 
in the former,\cite{c24} which qualitatively accords with its nonmetallic 
behavior and lower $T_{\rm C}$.

 For Sr$_{2}$MnMoO$_{6}$, its lattice constant is larger than SFMO
by $\approx$0.12 \AA~ and the Mn-O bond is quite long,\cite{c8} which indicates 
that the Mn
ion most probably takes a high-spin bivalence state (Mn$^{2+}$, $S$=5/2) with 
a large ionic radius. Another evidence to support this argument
is that there is no Jahn-Teller lattice distortion in 
Sr$_{2}$MnMoO$_{6}$\cite{c8} as expected 
for the $e_{g}$ half-filled Mn$^{3+}$ manganites. Owing to 
the Mo $4d$-O $2p$-Mn $3d$ hybridizations, however, the broadened 
up-spin Mn $e_{g}$ bands cross $E_{F}$ and thus become partly occupied, and so 
do the down-spin Mo$^{6+}$ $t_{2g}$ bands, as shown by the present LSDA FM 
calculation. For Sr$_{2}$CoMoO$_{6}$, 
the LSDA FM calculation gives a HM solution, which is caused by the 2/3 filled
down-spin Co$^{2+}$ $t_{2g}$ orbitals mixed with the O $2p_{\pi}$ and Mo$^{6+}$ 
$t_{2g}$ ones. Experimentally, the $c$-axis Co-O bond is a little elongated
in tetragonal Sr$_{2}$CoMoO$_{6}$ by 0.045 \AA~,\cite{c8} which lifts the 
triple-fold degeneracy of the Co $t_{2g}$
levels and splits them into a lower-energy doublet ($xz$/$yz$) and a higher
singlet ($xy$). While only the weak distortion seen by the Co $t_{2g}$ 
electrons, as shown by the LSDA calculation, is insufficient to open an 
insulating gap between the doublet and singlet. Instead the electron 
correlation determines the gap as seen below. 

By comparison of the down-spin Mo $t_{2g}$ level positions near $E_{F}$, it is 
suggested 
that Mo takes Mo$^{5+}$ in Sr$_{2}$CrMoO$_{6}$ and Sr$_{2}$FeMoO$_{6}$ but 
Mo$^{6+}$ in Sr$_{2}$MnMoO$_{6}$ and Sr$_{2}$CoMoO$_{6}$.
The Mo$^{6+}$ state in the latter two is also 
implied by the shrinked Mo-O 
bond length.\cite{c8} While the finite Mo component at $E_{F}$ in 
Sr$_{2}$MnMoO$_{6}$ and Sr$_{2}$CoMoO$_{6}$ is 
entirely ascribed to the $pdd$ hybridizations. 

As seen in Table I, however, the Mo $4d$ population varies insignificantly 
in Sr$_{2}$$M$MoO$_{6}$ ($M$=Cr,Mn,Fe,Co) due to strong intrinsic $pd$ 
covalency effects, and 
correspondingly, the $M$ $3d$ population increases by a step of $\sim$ 1, 
suggesting nearly same valence states of all $M$.
In addition, Sr and O retain the valence states of approximately +1.8 and --1.4
respectively, for all $M$. These results 
should be indicative of nearly a uniform valence state combination in 
Sr$_{2}$$M$MoO$_{6}$. Thus the present calculations support
a previous suggestion\cite{c8} that the $M$-independent spectral feature 
observed in Sr$_{2}$$M$MoO$_{6}$ at $\sim$2 eV
is ascribed to the O $2p$-Mo $4d$ transition. In addition, the relative 
spectral intensity
can be qualitatively explained in terms of the varying numbers of the empty 
Mo $t_{2g}$ states. It can be seen in Fig. 3 that the down-spin 
Mo $t_{2g}$ orbitals at $E_{F}$ have nearly the same partial filling in
Sr$_{2}$CrMoO$_{6}$ and Sr$_{2}$FeMoO$_{6}$, 
while they are almost unoccupied in Sr$_{2}$MnMoO$_{6}$ and Sr$_{2}$CoMoO$_{6}$.
Moreover, the number of the
Mo $t_{2g}$ holes is larger in Sr$_{2}$CoMoO$_{6}$ than in 
Sr$_{2}$MnMoO$_{6}$. These results account for 
nearly the same O $2p$-Mo $4d$ spectral feature of Sr$_{2}$CrMoO$_{6}$ and 
Sr$_{2}$FeMoO$_{6}$, the more intensive one 
of Sr$_{2}$MnMoO$_{6}$, and the most intensive one of 
Sr$_{2}$CoMoO$_{6}$.\cite{c8}  
  
 The LSDA+$U$ calculation causes nearly the same changes for 
SFMO as for BFRO and SFRO.
The Fe (Mo) spin moment is increased up to 4.23 (--0.60) $\mu_{B}$.   
The present values are larger than the lately reported ones of 3.97 
(--0.39) $\mu_{B}$,\cite{c10} and they are in good agreement with experimental 
ones of 4.2 (--0.5) $\mu_{B}$.\cite{c9} For Sr$_{2}$CrMoO$_{6}$, the up-spin 
Cr $t_{2g}$ bands
shift downward and scarcely contribute to $N(E_{F})$, and the Cr spin moment is
increased up to 2.42 $\mu_{B}$. The down-spin Mo $t_{2g}$ orbitals are
rather weakly hybridized  with the Cr $t_{2g}$ ones, and they form conduction
bands crossing $E_{F}$, which could be responsible for its relatively low 
room-temperature resistivity.\cite{c8} The present 
calculations
are not sufficient to reproduce the nonmetallic behavior of Ca$_{2}$FeReO$_{6}$
and Sr$_{2}$CrMoO$_{6}$, and therefore a better description of their 
nonmetallicity could resort to the site-disorder effects in them.
For Sr$_{2}$MnMoO$_{6}$, the up-spin Mn $e_{g}$ and down-spin Mo $t_{2g}$
bands contribute quite little to $N(E_{F})$ due to broadened bandwidths caused 
by the hybridization effects. The Mn spin moment of 4.54 $\mu_{B}$
indicates the high-spin Mn$^{2+}$ ($3d^{5}$) state with $S$=5/2, and a minor 
Mo spin moment of 0.04 $\mu_{B}$ suggests the formal Mo$^{6+}$ ($4d^{0}$) state.
This minor but positive Mo spin is induced by a little stronger
hybridization between the up-spin Mn $e_{g}$ and Mo $e_{g}$ orbitals. 
In terms of the Mn$^{2+}$/Mo$^{6+}$ model, an AFM ground state can be 
interpreted by superexchange interactions\cite{c25} via Mn-O-Mo-O-Mn 
bonds.\cite{c13} 
Correspondingly, the hybridized bandwidths shown in the above FM calculations
will be strongly suppressed in the AFM state, thus giving an insulating gap. 
Thus the AFM insulating ground state of Sr$_{2}$MnMoO$_{6}$ is 
reasonably explained. 
For Sr$_{2}$CoMoO$_{6}$, the electron correlation opens an insulating gap 
between the down-spin Mo $xz$/$yz$ doublet and $xy$ singlet. The Co spin moment
of 2.93 $\mu_{B}$ indicates a high-spin Co$^{2+}$ ($3d^{7}$,$S$=3/2) state. The
calculated small and negative Mo$^{6+}$ ($3d^{0}$) spin moment of 
--0.12 $\mu_{B}$ is ascribed to the stronger hybridization with the 
down-spin Co $xz$/$yz$ 
doublet. Like the case of Sr$_{2}$MnMoO$_{6}$, this Co$^{2+}$/Mo$^{6+}$ 
combination also leads to a superexchange-coupled AFM and insulating ground 
state of Sr$_{2}$CoMoO$_{6}$.   
 
\section*{IV. Conclusion} 

  A systematic electronic structure study of 
$A_{2}$FeReO$_{6}$ ($A$=Ba,Sr,Ca) and Sr$_{2}$$M$MoO$_{6}$ ($M$=Cr,Mn,Fe,Co) 
has been performed by
employing LSDA and LSDA+$U$ calculations. 

 (1) It is demonstrated that the HM 
character of BFRO, SFRO, and SFMO is not caused by the direct $M'$-$M'$ 
($M'$=Re or Mo) 
interactions but the indirect $M'$-O-Fe-O-$M'$
$pdd$-$\pi$ couplings which are simultaneously responsible for their FiM
character. 

(2) It is suggested that a finite $pdd$-$\sigma$ hybridized state being present at
$E_{F}$ in distorted CFRO contributes to its increasing $T_{\rm C}$ value,
compared with the case of BFRO. 

(3) The reduced $pdd$-$\pi$ coupling 
in CFRO and Sr$_{2}$CrMoO$_{6}$, as well as the site-disorder effect, could 
account for their nonmetallic behaviors. 

(4) The calculated level 
distributions and spin moments indicate the $M^{3+}$/Mo$^{5+}$ state for
Sr$_{2}$CrMoO$_{6}$ and Sr$_{2}$FeMoO$_{6}$ but the $M^{2+}$/Mo$^{6+}$ state 
for Sr$_{2}$MnMoO$_{6}$ and Sr$_{2}$CoMoO$_{6}$, in terms of
which both the itinerant FiM state of the former two and the superexchange-AFM 
insulating state of the latter two can be explained. 

(5) The population
analyses show that Sr$_{2}$$M$MoO$_{6}$ have nearly the same valence state 
combinations
due to strong intrinsic $pd$ covalency effects, which accounts for 
the similar O $2p$-Mo $4d$ spectral structures observed in them. \\
 
  The author thanks Y. Kakehashi and A. Yaresko for their discussions.
This work was cosponsored by Max Planck 
Society of Germany and National Natural Science Foundation of China.

\newpage

\begin{table} {Table I. $M$ $3d$ and $M'$ $4d$ or $5d$ population/spin moment
(in $\mu_{B}$) of $A_{2}$FeReO$_{6}$ ($A$=Ba,Sr,Ca) and Sr$_{2}$$M$MoO$_{6}$
($M$=Cr,Mn,Fe,Co) calculated by LSDA and LSDA+$U$.}
\begin{tabular} {rcccccccccccc}
 &&& \multicolumn{4}{c}{LSDA} &&& \multicolumn{4}{c}{LSDA+$U$} \\
 &&& $M$ &&& $M'$ &&& $M$ &&& $M'$ \\ \hline
$A$=Ba&&&5.76/3.72&&&4.46/--0.99&&&5.93/4.16&&&4.50/--1.22\\
Sr&&&5.62/3.75&&&4.36/--1.05&&&5.87/4.22&&&4.39/--1.27\\
Ca&&&5.56/3.76&&&4.14/--1.07&&&5.84/4.29&&&4.18/--1.33\\
$M$=Cr&&&3.48/1.67&&&3.34/--0.65&&&3.49/2.42&&&3.40/--0.53\\
Mn&&&4.62/3.94&&&3.51/--0.17&&&4.81/4.54&&&3.54/0.04\\
Fe&&&5.65/3.78&&&3.38/--0.44&&&5.89/4.23&&&3.43/--0.60\\
Co&&&6.85/2.64&&&3.51/--0.20&&&6.96/2.93&&&3.56/--0.12
\end{tabular}
\end{table}

\begin{figure}

\caption{ The LSDA and LSDA+$U$ DOS for the HM FiM state of $A_{2}$FeReO$_{6}$
($A$=Ba,Sr,Ca). The solid (dashed) line denotes the up (down) spin.
Fermi level is set at zero. \\}

\caption{ Charge density on the (001) plane of Ba$_{2}$FeReO$_{6}$,
with contour from 0.005 to 0.045 a.u.$^{-3}$ in a step of 0.005 a.u.$^{-3}$
(dashed line), and from 0.05 to 0.5 a.u.$^{-3}$ in a step of 0.05 a.u.$^{-3}$
(solid line). The Re ions are located at center and at corner, and the Fe ions
in the middle of edges. The Re and Fe ions are separated by oxygen. \\}

\caption{ The LSDA and LSDA+$U$ DOS for the FiM state of Sr$_{2}$$M$FeO$_{6}$
($M$=Cr,Fe) and for the assumed FM state of Sr$_{2}$$M$FeO$_{6}$
($M$=Mn,Co).  \\}

\end{figure}


\begin{references}
\bibitem{c24}  In the LSDA+$U$ calculation, the hybridization with the almost 
full-filled up-spin Cr $t_{2g}$ orbitals induces a finite positive Mo spin,
which partly counteracts the large negative Mo spin caused by the down-spin 
Mo $t_{2g}$ orbitals. The Mo $t_{2g}$ electrons are less itinerant in 
Sr$_{2}$CrMoO$_{6}$ than
in SFMO, which is also indicated by comparison of the down-spin Mo $t_{2g}$
bandwidths as seen in Fig. 3.
\end{references}
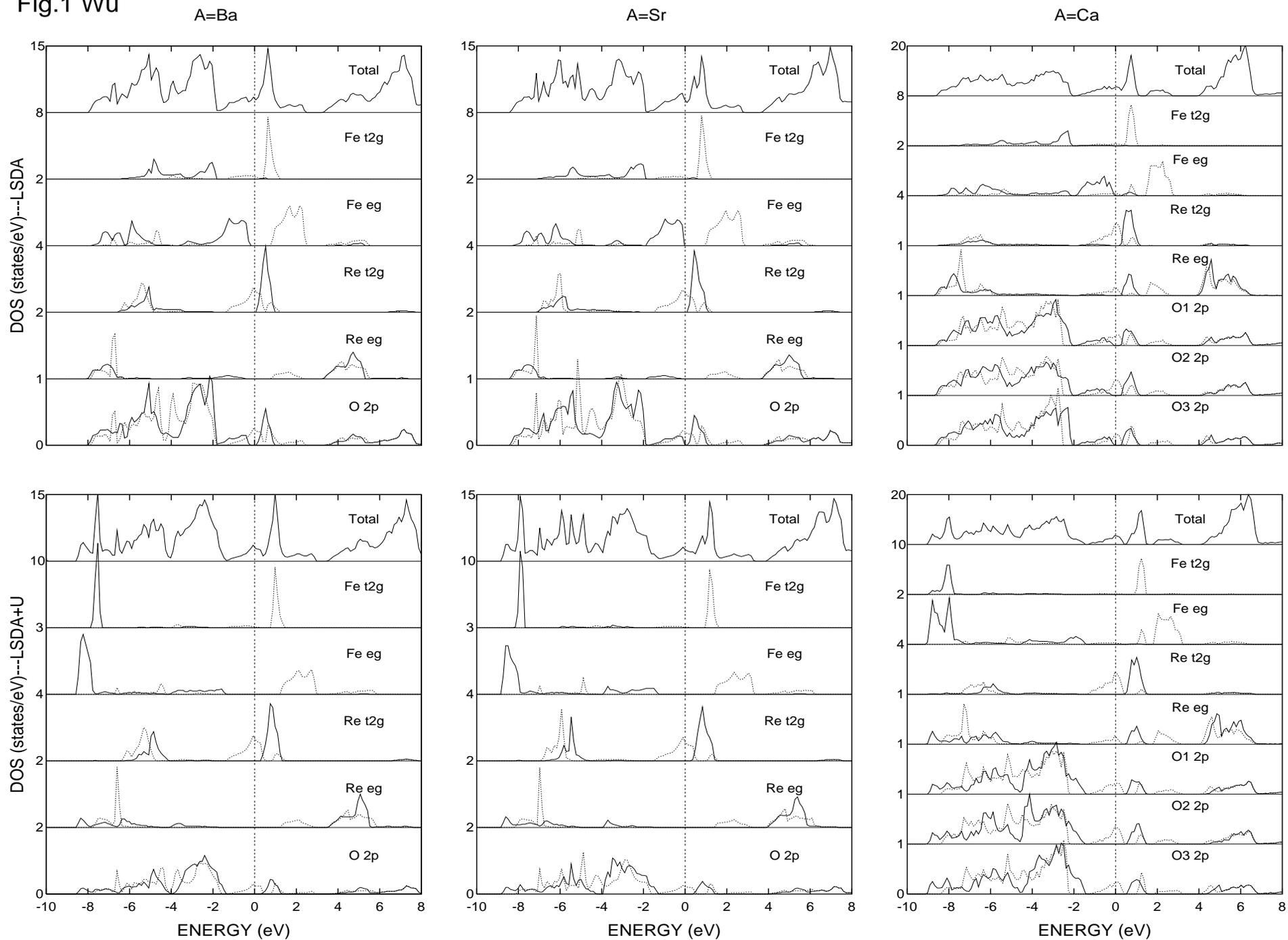
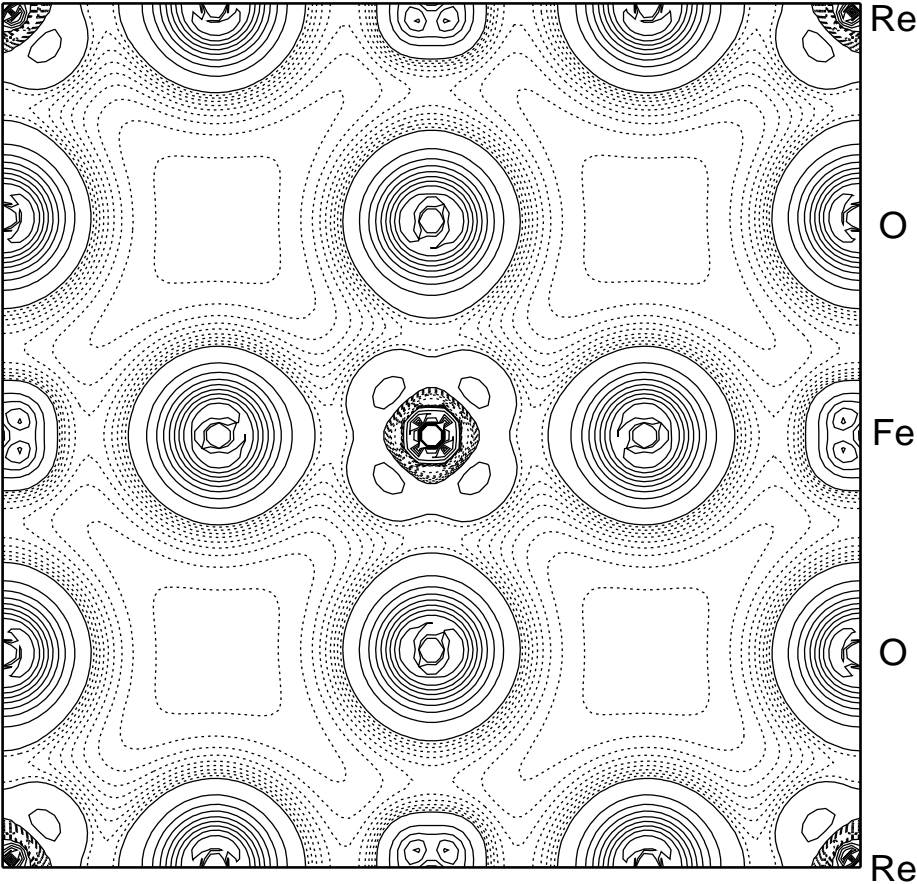
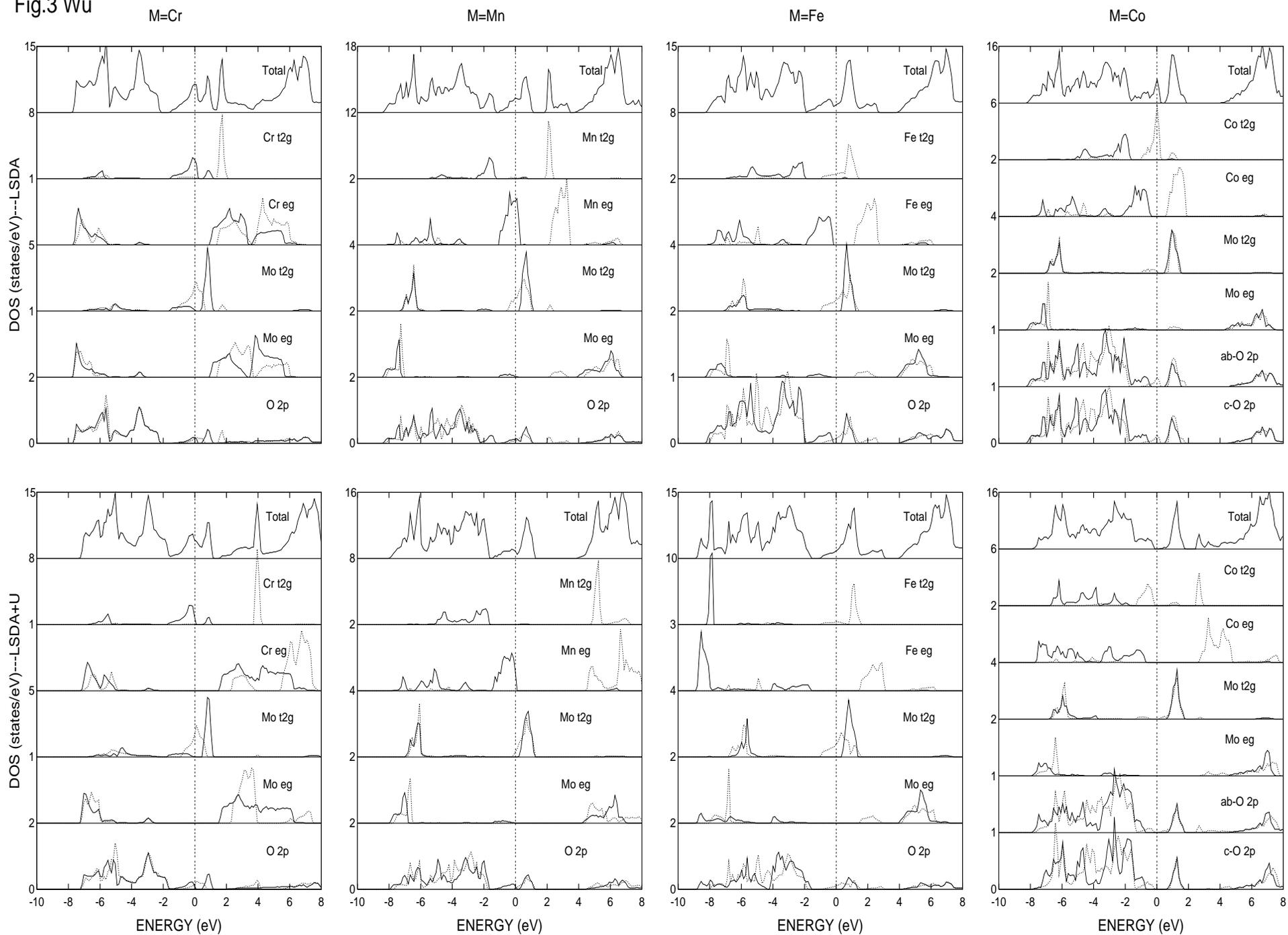
\end{document}